\hsize=160mm
\vsize=230mm
\hoffset=00mm
\voffset=00mm
\parskip=10pt
\parindent=15pt
\pageno=1
\headline={\ifnum\pageno>1 \hss \number\pageno\  \hss \else\hfill \fi}
\nopagenumbers
\vglue 12mm
\vskip 250pt
\centerline{\bf \ {ON THE CONSTRUCTION  OF $W_{_N}$- ALGEBRAS IN THE }}
\centerline{\bf \ {FORM OF $A_{_{N-1}}$- CASIMIR ALGEBRAS}}
\vskip 1mm
\centerline{\bf H.\ T.\ \"Ozer}
\vskip 1mm
\centerline{\it {Physics Department,\ Faculty of Science and Letters,}}
\centerline{\it {Istanbul Technical University,}}
\centerline{\it { 80626,\ Maslak,\ Istanbul,\ Turkey }}

\vskip 15mm
\noindent{\bf{Abstract}} \  Casimir W-algebras are shown to be exist
in such a way that the conformal spins of primary\ (generating) fields
coincide
with the orders of independent Casimir operators. We show here that
this coincidence can be extended further to the case that these
generating fields have the same eigenvalues with the Casimir operators.

\eject

\noindent{\bf{ 1. \ Introduction}}

Conformal invariance is of fundamental importance in two-di\-men\-si\-o\-nal
field theories and hence it founds remarkable applications in string theory [1]
and in the study of critical phenomena in statistical physics [2], as well as
in mathematics [3].\ Its
underlying symmetry algebra is Virasoro algebra which appears naturally
in two-dimensional field theories.\ The idea to extend Virasoro algebra with
the introduction of higher conformal spin generators is also seem to be
relevant in these theories.\ A seminal type of these extensions is the so-called
$W_{_N}$-algebras and Virasoro algebra is a $W_{_2}$-algebra within this framework.
$W_{_3}$ is,\ except superconformal algebras,\ one of the first non-trivial extensions
of $W_{_2}$  and it was constructed first by Zamolodchikov [4,5].\ There are several works
dealing with the classification and also the construction of this type
of algebras [6-10].\ To this end,\ a difficulty is the fact that,\ except $W_{_2}$ ,\ all the
extended ones are non-linear algebras.\ This poses great complications in
explicit constructions of $W_{_N}$ - algebras.\ One must emphasize here that very little
is known beyond $W_{_3}$ .\ There are, on the other hand,\ some recent efforts to
bring out a relation between these algebras and Casimir algebras[11,12]. \
The idea is principally based on the Sugawara construction
of $W_{_2}$-algebra [13].\ The purpose of our work is to establish a method in this
direction. For this,\ we made use of construction known as Miura
transformation [14-19]
with Feigin-Fuchs type of free massless scalar fields.\ It is seen that this
gives us the possibility to exploit a relation between $W_{_N}$- algebras  and
$A_{_{N-1}}$-Lie algebras.\
This brings us to the fact that one can define spin-2,3,4 primary fields of
$W_{_4}$-algebra [20] and also spin-5 primary field in such a way that their
eigenvalues are in one to one correspondence with 2,3,4 and 5 order
Casimir operator of $A_{_{N-1}}$ Lie algebras [21]. \

A definition of Casimir W-Algebras is given definitively in [20]
and also a relation is tried to be given with the eigenvalues of Casimir
operators in section 6 of the same paper,\ in examples only for second
and also third order Casimir operators.\ The studies for these orders are
also  given in [19].\  We tried to extend these studies
beyond third orders by calculating explicitly  the eigenvalues of
the generating fields and also the Casimir operators.

The paper is organized as follows.\
In section 2,
~we constructed $W_{_4}$-
algebra by utilizing a construction known as Miura transformation with
Feigin-Fuchs type of free massless scalar fields.\
In section 3,
~we show
our way of computing the eigenvalue spectrum of $W_{_N}$-algebras on highest
weight states of Fock space of $A_{_{N-1}}$ Lie algebras.\
In section 4,
~we obtain
a relation between  the eigenvalue spectrum of
$W_{_4}$-algebra by adding pure spin-5 conformal field and related
order Casimir eigenvalues for  irreducible representations of
$A_{_{N-1}}$Lie algebras.\
In appendix-A
we give an explicit form for our
pure spin-5
conformal field.\ All these have been possible with a dense application of
Mathematica Package program [22].

\noindent{\bf{ 2. \ The (quantum) Miura Transformation and $W_{_N}$-Algebras}}

The  $ W_N $-algebra is generated by a set of chiral currents $ \{ U_k (z)\} $ ,
~ of conformal dimension k~$~(k=1 , \cdots , N) $.\ Let us define a
differential  operator $ R_N (z) $ of degree N [19]
$$
R_{_N} (z)=
-\sum_{k=0}^N U_{_k} (z) (\alpha_0 \partial)^{N-k}=
: \prod_{j=1}^N \nabla_{_j}:,
\eqno(2.1)
$$
where $ \nabla_{_j} \equiv \alpha_{_0} \partial_{_z}- h_{_j}(z)$ and
the symbol $ : :$  shows the normal ordering of the fields $\varphi(z)$.\
Here  $ \varphi(z) $ is an $ N-1 $ component Feigin Fuchs-type of free
massless scalar fields .\ This transformation is called the (quantum)
Miura transformation and it determines completely the fields $\{U_{_k} (z)\}$
with
$$
h_{_j}(z)=i  \mu_{_j}\partial \varphi(z)
\eqno(2.2)
$$
Here,\ ${\mu}_i $'s, $(i=1 , \cdots , N) $ are the weights
of the fundamental representation of $ SU(N) $,~satisfying
$ \sum_{i=1}^N {\mu}_i=0 $ and ~${\mu}_i .{\mu}_j=\delta_{_{ij}}-{1\over N} $.
 \ The simple roots of
$ SU(N) $ are given by $ {\alpha}_i={\mu}_i-{\mu}_{i+1} $, $(i=1 , \cdots , N-1) $. \ The Weyl vector
of $ SU(N) $ is denoted as
$ \rho={{1} \over {2}} \sum_{\alpha > 0} {\bf \alpha^{+}} $ where $\alpha^{+}$ are
the positive roots of $ SU(N) $. \ A free scalar field $ \varphi (z) $
is a single-valued function on the complex plane and its mode expansion [19]
is given by
$$
i\,\partial{\varphi}(z)=\sum_{n \in Z} a_n z^{-n-1}.
\eqno(2.3)
$$
Canonical quantization gives the commutator relations
$$
[a_m,a_n]=m \delta_{_{m+n,0}}\ ,
\eqno(2.4)
$$
and these commutator relations are equivalent to the contraction
$$
 \partial{\varphi}(\underline{z)\partial{\varphi}(}w) =-{1 \over {(z-w)^2}} \ .
\eqno(2.5)
$$
By using single contraction $ \partial{\varphi}(\underline{z)\partial
{\varphi}(}w) $,\ a contraction of $ h_{_j}(z)$ with itself is given by [16]
$$
h_{_i}(\underline{z)\,h_{_j}(}w)={{\delta_{_{ij}}-{1\over N}}\over {(z-w)^2}}
\eqno(2.6)
$$
The fields $ \{ U_k (z)\} $ can be obtained by expanding $R_{_N} (z)$.\
We present  a first few one as in the following
$$
U_{_0}(z)=-1~,~~U_{_1}(z)=\sum_{_i}\,h_{_i}(z)=0~,~~
U_{_2}(z)=-\sum_{i<j}\,(h_{_i}\,h_{_j})(z)+\alpha_{_0}\,\sum_{_i}(i-1)\,
\partial{h}_{_i}(z)
\eqno(2.7)
$$
One can see that $ U_{_2} (z) \equiv T(z) $ has spin-2,~ which
is called the stress-energy tensor,~ $ U_k (z) $ has spin-k.
\ The standard OPE of $ T(z) $ with itself is
$$ T(z) T(w)={{c/2} \over (z-w)^4} +
{{2 T(w)} \over (z-w)^2}+{{\partial{T}(w)} \over {z-w}}+\cdots
\eqno(2.8)
$$
where the central charge,~for $ SU(N) $,~is given by
$$
c=(N-1)~(1-N(N+1) {\alpha_0}^2).
\eqno(2.9)
$$
A primary field $\phi_{_h}(z)$ with conformal spin-h must provide the following
OPE with $T(z)$
$$
T(z)\phi_{_h} (w)={h\,\phi_{_h} (w) \over {(z-w)}^2}
+{{\partial\,\phi_{_h} (w)} \over {(z-w)}}+\cdots
\eqno(2.10)
$$
Therefore the fields $ \{ U_k (z) \} $ are not primary because
$$
\eqalign{
T(z)U_{_k}(w)&={1 \over 2}\!\sum_{s=1}^k {(N\!-\!k\!+\!s)! \over (N\!-\!k)!}
a_{_0}^{s\!-\!2}\Big(((s\!-\!1)(N\!-\!1)+2(k\!-\!1))a_{_0}^2\!-\!{{s\!-\!1}
\over N}\Big) {{U_{_{k\!-\!s}}(w)} \over (z\!-\!w)^{s\!+\!2}}\cr
&~~~~~~~+{{k U_{_k}(w)} \over (z-w)^2}+{{\partial{U_{_k}(w)}} \over
(z-w)}+(TU_k)(z)+\cdots \cr}
\eqno(2.11)
$$
Using above OPE,\ we want to construct $W_{_4}$-algebra.\ Therefore
we must obtain spin-3 and spin-4 primary fields. \ Here we first write
down the spin-3 primary field for $ SU(N) $ as
$$
\overline{U}_{_3}(z) =U_3 (z)-{(N-2) \over 2}~ \alpha_{_0}\partial{T}(z)
\eqno(2.12)
$$
and the spin-4 primary field as
$$
\overline{U}_{_4} (z)= U_{4} (z)-{(N-3) \over 2}\,a_{_0}\,~\partial U_{3} (z)
+{(N-2)(N-3) \over {4N(22+5c)}}\,[-3+N(13+3N+2c)\,a_{_0}^2]\,~\partial^2 T(z)
$$
$$
+{(N-2)(N-3) \over {2N(22+5c)}}\,[5-N(5N+7)\,a_{_0}^2]\,~(TT)(z)
\eqno(2.13)
$$

To obtain  OPE of the two primary fields $\{\overline{U}_{_k} (z)\}$ and
$\{\overline{U}_{_k \prime} (z)\}$ which
gives the central terms in the known form,~ we must take care of the normalized
forms of all the primary fields $\{\overline{U}_{_k} (z)\}$.\ Therefore the
normalized form of the $W_{_4}$-algebra generators are given by the following
expressions
$$
\overline{U}_{_3}(z)    =\sqrt{\theta_{_W}} W(z) ~~,
~~~ \overline{U}_{_4} (z)    =\sqrt{\theta_{_L}} L(z)
\eqno(2.14)
$$
where $ \theta_{_W} $ and $ \theta_{L}$ are the normalization
factors for $ SU(4) $ and their explicit form are given
$$
\theta_{_W}={{c+7} \over {10}}~~,
~~ \theta_{L}={{(114+7c)(c+7)(c+2)} \over {300(22+5c)}}
\eqno(2.15)
$$
respectively.\ These results are being in line with those of ref.[9,20].\ On
the other hand, we must emphasize here that the stress-energy
tensor $T(z)$ is not given in the normalized form.\ A straightforward calculation
gives us the first non-trivial OPE of $ W(z) $ with itself for $ SU(4) $
$$
W(z)\,W(w)={{c/3} \over {(z-w)}^6}+{2\,T(w) \over {(z-w)}^4}
+{{\partial\,T(w)} \over {(z-w)}^3}
+{1 \over {(z-w)}^2}[2\,\beta_{_\Lambda}\Lambda(w)+{3 \over
10}\,\partial{T}(w)+\beta_{_L}\,L(w)]
$$
$$
+{1 \over {(z-w)}}[\beta_{_\Lambda}\partial{\Lambda}(w)+{1 \over
15}\,{\partial}^3\,T(w)+{\beta_{_L}\over 2}\,\partial{L}(w)] +(WW)(w)+\cdots
\eqno(2.16)
$$
$$
\beta_{_\Lambda}={16 \over {22+5c}}~,
~\beta_{_L}={{4\sqrt{\theta_{_L}}} \over {\theta_{_W}}}
=\sqrt{ {16 \over 3} {{(114+7c)(c+2)} \over {(c+7)(22+5c)}}}
\eqno(2.17)
$$
Previously, we have written the normalization factor $\theta_{_W}$
for  $SU(4)$ in equation (2.15). In addition to this,
after some calculations it is also possible to write $\theta_{_W}$
for  $SU(N)$ in general form
$$
\theta_{_W}={{(N-2)(-2+2c-N+cN+3N^2)} \over {2(N-1)N(N+1)}}.
\eqno(2.18)
$$
The second non-trivial OPE of $W(z)$ with $L(w)$ takes the form
$$
W(z)\,L(w)=\eta_{_W}\Big[{W(w) \over {(z-w)}^4}
\!+\!{1 \over 3}{{\partial{W}(w)} \over {(z\!-\!w)}^3}\Big]
+\!{1 \over {(z\!-\!w)}^2}\Big[\eta_{_{TW}}\Big((TW)(w)\!-\!{3 \over
10}{\partial}^2\,W(w)\Big)\!+\!{\eta_{_W}\over {14}}\,{\partial}^2 W(w)\Big]
$$
$$
+\!{1 \over {(z\!-\!w)}}\Big[\eta_{_{\partial{(TW)}}}\partial\Big((TW)(w)
\!-\!{3 \over 10}{\partial}^2 W(w)\Big)
-\!\eta_{_{T\partial{W}}}\Big((T\partial{W})(w)\!-\!{{c\!+\!17} \over 84}
{\partial}^3 W(w)\Big)\Big]\!+\!(WL)(w)\!+\!\cdots
\eqno(2.19)
$$
where
$$
\eta_{_W}^2={3 \over
4}\,\beta_{_L}~~,~~\eta_{_{TW}}={39
\over {114+7c}}\,\beta_{_L}~~,~~
\eta_{_{\partial{(TW)}}}={15 \over 4}\,{(22+5c) \over
{(c+2)(114+7c)}}\,\beta_{_L}~~,~~\eta_{_{T\,\partial{W}}}={3 \over
{4(c+2)}}\,\beta_{_L}
\eqno(2.20)
$$
Finally,~the last non-trivial OPE of $L(z)$ with itself
takes the form
$$
L(z)L(w)\!=\!{{c/4} \over {(z\!-\!w)}^8}\!+\!{2T(w) \over {(z\!-\!w)}^6}
\!+\!{{\partial{T}(w)} \over {(z\!-\!w)}^5}
+\!{1 \over {(z\!-\!w)}^4}[{3 \over
10}{\partial}^2{T}(w)\!+\!2\rho_{_\Lambda}\Lambda(w)\!-\!
\rho_{_L}L(w)]
$$
$$
+\!{1 \over {(z\!-\!w)}^3}[{1 \over 15}{\partial}^3T(w)\!+\!\rho_{_\Lambda}
\partial{\Lambda}(w)\!-\!{{\rho_{_\Lambda}} \over 2}\partial{L}(w)]
+\!{1 \over {(z\!-\!w)}^2}[\rho_{_{{\partial}^2L}}{\partial}^2L(w)\!+\!
\rho_{_{T{\partial}^2T}}(T{\partial}^2T)(w)\!+\!
\rho_{_{\partial{T}\partial{T}}}(\partial{T}\partial{T})(w)
$$
$$
+\!\rho_{_{(TT)T}}((TT)T)(w)
+\!\rho_{_{WW}}(WW)(w)\!+\!\rho_{_{LT}}(LT)(w)\!+\!
\rho_{_{{\partial}^2{\Lambda}}}{\partial}^2{\Lambda}(w)]
$$
$$
+\!{1 \over {(z\!-\!w)}}[\rho_{_{{\partial}^3L}}
{\partial}^3L(w)\!+\!\rho_{_{\partial(T{\partial}^2T)}}\partial(T{\partial}^2T)(w)
\!+\!\rho_{_{\partial(\partial{T}\partial{T})}}\partial(\partial{T}\partial{T})(w)
+\!{1 \over 2}\rho_{_{(TT)T}}\partial{((TT)T)}(w)
$$
$$
\!+\!{\rho_{_{WW}} \over 2}\partial{(WW)}(w)\!+\!{1 \over 2}
\rho_{_{LT}}\partial{(LT)}(w)
+\!\rho_{_{{\partial}^3{\Lambda}}}{\partial}^3{\Lambda}(w)]\!+\!(LL)(w)\!+\!\cdots
\eqno(2.21)
$$
where
$$
\rho_{_{\Lambda}}\!=\!{21 \over
{22+5c}}~~,~~
\rho_{_L}\!=\!\sqrt{{{27(c^2+c+218)^2}
\over {(7c+114)(5c+22)(c+7)(c+2)}}}
$$
$$
\rho_{_{{\partial}^2{L}}}\!=\!{{4194\!+\!137c\!-\!5c^2} \over
{\sqrt{48(7c\!+\!114)(5c\!+\!22)(c\!+\!7)(c\!+\!2)}}}~~,~~
\rho_{_{T{\partial}^2{T}}}\!=\!{{-1596\!-\!2068c\!+\!25c^2} \over
{70(c\!+\!2)(114\!+\!7c)}}
$$
$$
\rho_{_{\partial{T}\partial{T}}}\!=\!{{1806\!-\!907c\!+\!10c^2} \over
{28(7c\!+\!114)(c\!+\!2)}}~~,~~
\rho_{_{(TT)T}}\!=\!{{96(9c\!-\!2)} \over
{(7c\!+\!114)(5c\!+\!22)(c\!+\!2)}}~~,~~
\rho_{_{WW}}\!=\!{{45(22\!+\!5c)} \over
{2(7c\!+\!114)(c\!+\!2)}}
$$
$$
\rho_{_{LT}}\!=\!{{48(114\!+\!7c)} \over
{(5c\!+\!22)(c\!+\!7)(c\!+\!2)}}~~,~~
\rho_{_{{\partial}^2{\Lambda}}}\!=\!{{-(23016\!+\!12948c\!-\!3190c^2\!+\!25c^3)
} \over {28(7c\!+\!114)(5c\!+\!22)(c\!+\!2)}}
$$
$$
\rho_{_{{\partial}^3{L}}}\!=\!{{2424\!+\!70c\!-\!c^2} \over
{\sqrt{48(7c\!+\!114)(5c\!+\!22)(c\!+\!7)(c\!+\!2)}}}~~,~~
\rho_{_{\partial(T{\partial}^2{T})}}\!=\!{{3(-1444\!-\!1592c\!+\!5c^2)} \over
{280(7c\!+\!114)(c\!+\!2)}}
$$
$$
\rho_{_{\partial(\partial{T}\partial{T})}}\!=\!{{3(526\!-\!345c\!+\!c^2)} \over
{56(7c\!+\!114)(c\!+\!2)}}~~,~~
\rho_{_{{\partial}^3{\Lambda}}}\!=\!{{-3(28568\!+\!15676c\!-\!1934c^2\!+\!5c^3)
} \over {112(7c\!+\!114)(5c\!+\!22)(c\!+\!2)}}
\eqno(2.22)
$$
These are also the same as in ref.[9].

\noindent{\bf{ 3.\ The Eigenvalue Spectrum of $W_{_N}$-Algebras }}

We denote the Fock space of a free massless scalar field
$h_{_j}(z)=i\,\mu_{_j}\partial{\varphi}(z)$ in (2.2)  by  $F_{\Lambda}$,
~where  $\Lambda$ is the dominant weight of the Lie algebra  $A_{N-1}$,
~which can be expressed as
$
\Lambda=\sum_{i=1}^{N-1}\,r_{i}\,\lambda_{i}
$
where  $\lambda_i$ 's are   fundamental dominant weights which are defined by
$\lambda_i~.~\alpha_{_j}=\delta_{ij}$ as dual vectors to simple roots
$\alpha_{_i}$.\ Cartan matrices $C_{ij}$ are then defined by
$C_{ij}=\alpha_{_i}\,.\,\alpha_{_j}$ being in accordance with
the Dynkin diagrams
$${o--o--o--o\cdot\cdot\cdot o--o--o}$$
$$\leftline{~~~~~~~~~~~~~~~~~~~~~~~~~~~~~~~~~~~~~~~~~~~~~$_{_{1}}
$~~~~~~$_{_{2}}$~~~~~~$_{_{3}}$~~~~~~ .~.~. ~~~~~$_{_{N-1}}
$~~~~~~$_{_{N}}$}
\eqno(3.1)
$$
\noindent of $ A_{N-1}$ chain where $N=1,2,3,\cdot\cdot\cdot$.~ The parameters
$r_i$  are taken to be positive integers including zero,~and also
$\Lambda$ labels the eigenvalue of the scalar zero modes  $a_{0}^j=p^j $ on the
Fock space vacuum  $\vert \Lambda>$ [19].~This can be written as
$$
a_{0}^j \vert \Lambda>=\Lambda \vert \Lambda>.
\eqno(3.2)
$$

The eigenvalues $\{U_k(\Lambda)\}$ of the zero mode of
$\{U_k(z)\}$  on the highest weight states of Fock space $F_{\Lambda}$:
are given by [19]
$$
U_k(\Lambda)=(-1)^{k-1}\sum_{i_1<i_2<...<i_k}\,\prod_{j=1}^k\Big[(\Lambda,
\mu_{i_{_j}})+(k-j)\,\alpha_{_0} \Big]
\eqno(3.3)
$$

We give a general definition to the $\Theta_{(n_1,n_2,...,n_k)}$,
$$
\Theta_{(n_1,n_2,...,n_k)}=\sum_{i_1<i_2<...<i_k}^N\,\Big\{\prod_{j=1}^k\,\theta_
{i_{_j}}^{n_{_j}}\Big\}
\eqno(3.4)
$$
where $ \theta_i=(\Lambda+\alpha_{_0}\,\rho~,~\mu_i)$
and $ \Theta_{(1)}=\sum_{i=1}^N\,\theta_i=0$  since
$ \sum_{i=1}^N\,\mu_i=0$.\ Therefore,~we give the results
$k=2,3,4$ and $5$ respectively
$$
U_{1}(\Lambda)=\Theta_{(1)}=0~~,~~
U_{2}(\Lambda)=-\Theta_{(1,1)}-{1 \over 4}\,
\pmatrix{N+1  \cr   3 }
\,\alpha_{_0}^2
\eqno(3.5)
$$
$$
U_{3}(\Lambda)=\Theta_{(1,1,1)}+(N-2)\,\alpha_{_0}\,\Theta_{(1,1)}+
\pmatrix{N+1  \cr   4  }
\alpha_{_0}^3
\eqno(3.6)
$$
$$
U_{4}(\Lambda)=-\Theta_{(1,1,1,1)}-{3 \over
2}\,(N-3)\,\alpha_{_0}\,\Theta_{(1,1,1)}
$$
$$
-{1 \over24}\,(N-2)(N-3)(N+23)\,\alpha_{_0}^2\,\Theta_{(1,1)}-{(223+5N) \over
48}\,
\pmatrix{N+1  \cr 5 }
\,\alpha_{_0}^4
\eqno(3.7)
$$
and
$$
U_{5}(\Lambda)=\Theta_{(1,1,1,1,1)}+2(N-4)\,\alpha_{_0}\,\Theta_{(1,1,1,1)}
\!\!+\!\!{{(N\!\!-\!\!3)(N\!\!-\!\!4)(N\!\!+\!\!43)} \over
24}\alpha_{_0}^2\Theta_{(1,1,1)}
$$
$$
\!\!+\!\!{1 \over
12}(N\!\!-\!\!2)(N\!\!-\!\!3)(N\!\!-\!\!4)(N\!\!+\!\!11)\alpha_{_0}^3\Theta_{(1
,1)}
+{(103+5N) \over 4}\,
\pmatrix{N+1  \cr 6 }
\,\alpha_{_0}^5
\eqno(3.8)
$$
where
$$
\Theta_{(1)}=0~~,~~
\Theta_{(1,1)}=-{1 \over 2}\,\Theta_{(2)}~~,~~
\Theta_{(1,1,1)}={1 \over 3}\,\Theta_{(3)}
$$
$$
\Theta_{(1,1,1,1)}=-{1 \over 4}\,\Theta_{(4)}+{1 \over 8}\,\Theta_{(2)}^2~~,~~
\Theta_{(1,1,1,1,1)}={1 \over 5}\,\Theta_{(5)}-{1 \over 6}\,\Theta_{(2)}
\Theta_{(3)}
\eqno(3.9)
$$
After some calculations it can be shown that the eigenvalues of the primary
fields are $U_{_2}(z)\equiv T(z)\equiv \overline{U}_{_2}(z)$, spin-3 $\overline{U}_{_3}(z)$,
spin-4 $\overline{U}_{_4}(z)$ and spin-5 $\overline{U}_{_5}(z)$. As an example, the
eigenvalue of primary field spin-5 $\overline{U}_{_5}(z)$ is calculated in
appendix-A.
Finally, we now write down all the eigenvalues of the primary fields
in the following form:
$$
\overline{U}_{_2}(\Lambda)={1 \over 2}\,\Theta_{(2)}-{1 \over 4}\,
\pmatrix{N+1  \cr
3 \cr}\,\alpha_{_0}^2~~,~~
\overline{U}_{_3}(\Lambda)={1 \over 3}\,\Theta_{(3)}
\eqno(3.10)
$$
$$
\overline{U}_{_4}(\Lambda)={1 \over 4}\Theta_{(4)}+{3 \over
{4N(22+5c)}}[5\!\!-7N\!\!+N(3N^2\!\!-\!\!7)\alpha_{_0}^2]\Theta_{(2)}^2
$$
$$
-{{(N-2)(N-3)} \over
{4N(22+5c)}}[9\!\!-N(11N\!\!+\!\!15)\alpha_{_0}^2]\Theta_{(2)}
+{{(N-1)(N-2)(N-3)(N+1)} \over
{240(22+5c)}}[45\!\!-N(50N\!\!+\!\!69)\alpha_{_0}^2]
\eqno(3.11)
$$
and
$$
\overline{U}_{_5}(\Lambda)={1 \over 5}\,\Theta_{(5)}
+{1 \over
{720N(114+7c)}}[10080+N(7N^4-71N^3+173N^2+191N-19500)
$$
$$
-N^2(N-1)(N+1)(7N^3-36N^2-7N-684)\alpha_{_0}^2]\Theta_{(2)}\Theta_{(3)}
$$
$$
+{{(N-3)(N-4)} \over
{1036800N(114+7c)}}[(N-5)(N-1)^2
N^3(7N+13)(N^2+5N+24)\alpha_{_0}^4
$$
$$
-2N^2(N-1)(N+1)(7N^5-22N^4-72N^3-398N^2-31735N-56580)\alpha_{_0}^2
$$
$$
+7N^7-57N^6+38N^5+382N^4-66285N^3+209915N^2
+604800N-6220800]\Theta_{(3)}
$$
$$
\eqno(3.12)
$$
\noindent{\bf{ 4.\ A Relation Between Casimir Eigenvalues of $A_{_{N-1}}$-Lie
Al\-geb\-ras and Eigenvalue Spectrum of $W_{_N}$-Algebras}}

In this section,\ we will try to establish a relation between Casimir
eigenvalues of $A_{N-1} $ Lie algebras  and the eigenvalues of zero modes
of generating fields.\ Let $\Lambda$ be a dominant weight for $A_{N-1}$ Lie
algebra.\ For an irreducible  representation $Rep[\Lambda]$,\ the eigenvalues
of a Casimir operator of order N  are given in ref.\ [21] as
in the following form
$$
C_{_N}[Rep[\Lambda]]={\rm dim}\,Rep[\Lambda]\,P_{_N}\,[Rep[\Lambda]]
\eqno(4.1)
$$
where the polinomial  $P_{_{N}}\,[Rep[\Lambda]]$ is a $N$ order polinomial of
$\theta_{_i}\,~(i=1,2,...,N)$.\ In the following we will
give explicit expression of the eigenvalues for the Casimir operators

$$
C_{_2}[Rep[\Lambda]]=-{\rm dim}\,Rep[\Lambda]\Big({12 \over
{(N-1)N(N+1)}}\,\Theta_{(2)}-1\Big)
\eqno(4.2)
$$
$$
C_{_3}[Rep[\Lambda]]={\rm dim}\,Rep[\Lambda]\,\Theta_{(3)}
\eqno(4.3)
$$
$$
C_{_4}[Rep[\Lambda]]={\rm
dim}\,Rep[\Lambda](\alpha_{_1} P_{_4}\,[Rep[\Lambda]]+\alpha_{_2}
P_{_{2,2}}\,[Rep[\Lambda]])
\eqno(4.4)
$$
and
$$
C_{_5}[Rep[\Lambda]]={\rm
dim}\,Rep[\Lambda](\beta_{_1} P_{_5}[Rep[\Lambda]]+\beta_{_2}
P_{_{3,2}}\,[Rep[\Lambda]])
\eqno(4.5)
$$
where
$$
P_{_4}\,[Rep[\Lambda]]={{720 N [(N^2+1)\Theta_{(4)}+(3-2N^2)\Theta_{(2)}^2]} \over
{(N-3)(N-2)(N-1)N^2(N+1)(N+2)(N+3)}} \ +\ 1
\eqno(4.6)
$$
$$
P_{_{2,2}}[Rep[\Lambda]]\!\!=\!\!{{720[2N(2N^2\!-\!3)\Theta_{(4)}\!-\!
(N^4\!-\!6N^2\!+\!18)\Theta_{(2)}^2]}
\over {(N\!-\!3)(N\!-\!2)(N\!-\!1)N^2(N\!+\!1)(N\!+\!2)(N\!+\!3)(6\!+\!5N^2)}}
+{120N \over {(N-1)(N+1)(6+5N^2)}}\Theta_{(2)}-1
\eqno(4.7)
$$
$$
P_{_5}\,[Rep[\Lambda]]={{N(5+N^2)} \over
{5(2-N^2)}}\Theta_{(5)}+\Theta_{(3)}\Theta_{(2)}
\eqno(4.8)
$$
and
$$
P_{_{3,2}}[Rep[\Lambda]]\!\!=\!\!{{72 N(N^2\!-\!2)\Theta_{(5)}\!-\!
12(N^4+24)\Theta_{(3)}\Theta_{(2)}} \over
{(N\!-\!4)(N\!-\!3) N^3 (N\!+\!3)(N\!+\!4)}}\!+\!\Theta_{(3)}
\eqno(4.9)
$$
with some arbitrary constants $\alpha_i$ and $\beta_i$
\par Finally, we will show that there is a relation between the eigenvalues of
the primary fields and $P_{N}\,[Rep[\Lambda]]$ polinomials.~ In other words,~
it is possible that the eigenvalues of the primary fields can be written with
respect to $P_{N}\,[Rep[\Lambda]]$,~and the resulting expressions will be
as in following:
$$
\overline{U}_{_2}(\Lambda)={{(N-1)N(N+1)} \over 24} \Big(P_{_2}[Rep[\Lambda]]-1\Big)
\eqno(4.10)
$$
$$
\overline{U}_{_3}(\Lambda)={1 \over 3}P_{_3}[Rep[\Lambda]]
\eqno(4.11)
$$
$$
\overline{U}_{_4}(\Lambda)=\gamma_{_1}P_{_4}[Rep[\Lambda]]+\gamma_{_2}P_{_{2,2}}
[Rep[\Lambda]] +\gamma_{_3}P_{_2}[Rep[\Lambda]]+\gamma_{_4}
\eqno(4.12)
$$
and
$$
\overline{U}_{_5}(\Lambda)=\sigma_{_1}P_{_5}[Rep[\Lambda]]+\sigma_{_2}P_{_{3,2}}
[Rep[\Lambda]] +\sigma_{_3}P_{_3}[Rep[\Lambda]]
\eqno(4.13)
$$
where
$$
\eqalign{
\gamma_{_1}&\!\!=\!\!{{(N\!\!-\!\!1)(N\!\!-\!\!2)(N\!\!-\!\!3)(N\!\!+\!\!1)} \over
{2880(22\!\!+\!\!5c)N}}\,[36
\!+\!N(5N^2\!\!+\!\!42N\!\!+\!\!66)\!\!+\!\!N(N\!\!+\!\!2)(N\!\!+\!\!3)
(6\!\!-\!\!5N^2)\alpha_{_0}^2] \cr
\gamma_{_2}&\!\!=\!\!{{(N\!\!-\!\!1)(N\!\!-\!\!2)(N\!\!-\!\!3)(N\!\!+\!\!1)(6+5N^2)}
\over {2880(22\!\!+\!\!5c)N}}\,[6
\!+\!11N\!\!+\!\!N(N\!+\!\!2)(N\!\!+\!\!3)
\alpha_{_0}^2] \cr
\gamma_{_3}&\!\!=\!\!{{(N\!\!-\!\!1)(N\!\!-\!\!2)(N\!\!-\!\!3)(N\!\!+\!\!1)} \over
{288(22\!\!+\!\!5c)N}}\,[54
\!+\!6N\!\!+\!\!11N^2\!\!+\!\!N(N^3\!\!+\!\!5N^2\!\!-\!\!60N\!\!-\!\!90)
\alpha_{_0}^2] \cr
\gamma_{_4}&\!\!=\!\!{{N(N\!\!-\!\!1)(N\!\!-\!\!2)(N\!\!-\!\!3)(N\!\!+\!\!1)(6+5N)
(\alpha_{_0}^2-1)} \over {240(22\!\!+\!\!5c)}} \cr
\sigma_{_1}&\!\!=\!\!{(N^2\!\!-\!\!2) \over
{24(114\!\!+\!\!7c)N^3(N\!\!+\!\!3)(N\!\!+\!\!4)}}[-3456\!\!-\!\!5602N\!\!-\!\!
3788N^2\!\!-\!\!89N^3 \cr
&\!\!+\!\!22N^4\!\!-\!\!7N^5\!\!+\!\!(N^3\!\!-\!\!N)(336\!\!+\!\!310N\!\!+\!\!154N^2\!\!+\!\!13N^3\!\!+\!\!7N^4)\alpha_{_0}^2] \cr
\sigma_{_2}&\!\!=\!\!{{(N\!\!-\!\!3)(N\!\!-\!\!4)} \over
{8640(114\!\!+\!\!7c)}}
[8640\!\!+\!\!14005N\!\!+\!\!110N^2\!\!+\!\!30N^3\!\!+\!\!22N^4\!\!-\!\!7N^5]\cr
&\!\!+\!\!(N^3\!\!-\!\!N)(-840\!\!-\!\!775N\!\!+\!\!35N^2\!\!+\!\!13N^3\!\!+\!\!7N^4)\alpha_{_0}^2] \cr
\sigma_{_3}&\!\!=\!\!{{(N\!\!-\!\!3)(N\!\!-\!\!4)} \over
{1036800(114\!\!+\!\!7c)N}}[\!-6220800\!\!-\!\!432000N\!\!-\!\!1470685N^2\!\!-\!
\!79485N^3 \cr
&\!\!-\!\!3218N^4
\!\!-\!\!2602N^5\!\!+\!\!2N(N^3\!\!-\!\!N)(106980\!\!+\!\!78235N\!\!-\!\!1702N^2\!\!-\!\!
708N^3 \cr
&\!\!-\!\!398N^4\!\!-\!\!7N^5)\alpha_{_0}^2
\!\!+\!\!(N\!\!-\!\!5)(N\!\!-\!\!1)^2 N^3 (13\!\!+\!\!7N)(24\!\!+\!\!5N\!\!+
\!\!N^2)\alpha_{_0}^4]
\cr}
\eqno(4.14)
$$

\noindent{\bf{ACKNOWLEDGEMENT}} \ I would like  to thank H. R. Karaday\i
~for his valuable discussions and excellent guidance  throughout this research.
\vskip 5mm

\leftline{\bf \ {APPENDIX-A}}

The primary field $\overline{U}_{_5}(z)$,~which was used in (3.12) with
eigenvalue $\overline{U}_{_5}(\Lambda)$,~is in the form of
$$
\overline{U}_{_5}(z)= U_{_5}(z)-{(N-4) \over 2}\,a_{_0}\,\partial{U_{_4}}(z) \ +
$$
$$
{3 \over 4}\,{ (N-3)(N-4) \over {N(114+7c)}}\,[-2+N(20+C+2N)\,a_{_0}^2]\, \partial^{2} {U_{_3}}(z) \ +
$$
$$
{(N-2)(N-3)(N-4)\,a_{_0} \over {12N(114+7c)}}\,[9-N(33+C+9N)\,a_{_0}^2]\, \partial^{3} {U_{_2}}(z) \ +
$$
$$
{(N-3)(N-4) \over {N(114+7c)}}\,[7-N(13+7N)\,a_{_0}^2]\, (U_{_2}U_{_3})(z) \ +
$$
$$
{ (N-2)(N-3)(N-4)\,a_{_0} \over {2N(114+7c)}}\,[-7+N(13+7N)
\,a_{_0}^2]\, (U_{_2} \partial{U_{_2}})(z)
\eqno(A.1)
$$
\vskip 5mm
\leftline{\bf REFERENCES}
\vskip 5mm
\noindent 1. M.B. Green, J.H. Schwarz and E. Witten, Superstring Theory ,
Vols. 1, 2 (Cambridge Univ. Press. Cambridge. 1987).

\noindent 2. C. Itzykson, H. Saluer and J.B. Zuber, eds, Conformal
Invariance and Applications to Statistical Mechanics
(World Scientific, Singapore, 1988).

\noindent 3. A.A. Belavin, A.M. Polyakov and A.B. Zamolodchikov,
Nucl. Phys. B241 (1984) 333.

\noindent 4. A.B. Zamolodchikov, Theor. Math. Phys. 65 (1985) 205.

\noindent 5. V.A. Fateev and A.B. Zamolodchikov, Nucl. Phys. B280 (1987) 644.

\noindent 6. K. Hamada and M. Takao, Phys. Lett. B209 (1988) 247.

\noindent 7. D.H. Zhang, Phys. Lett. B232 (1989) 323.

\noindent 8. H.G. Kausch and G.M.T. Watts, Nucl. Phys. B354 (1991) 740.

\noindent 9. Blumenhagen et al., Nucl. Phys. B361 (1991) 255.

\noindent 10. C. Hansoy, A General Method On the Construction of
$ W_{_N}$-Al\-geb\-ras, MS Thesis, Istanbul Technical University, 1993.

\noindent 11. J. Thierry-Mieg, Generalizations of the Sugawara construction,
in : Nonperturbative Quantum Field Theory, eds G.\ 't Hooft et al., Proc.
Cargese School 1987 (Plenum Press, New York, 1988) p.567.

\noindent 12. F. Bais, P. Bouwknegt, M. Surridge and K. Schou\-tens, Nucl. Phys.
B304 (1988) 348; 371.

\noindent 13. H.\ Sugawara,\ A field theory of currents,
\ Phys.\ Rev. \ 170\ (1968)\ 1659.

\noindent 14. V.G. Drinfel'd and V. Sokolov, J. Sov. Math. 30 (1985) 1975.

\noindent 15. V.A. Fateev and S.L. Lukyanov, Int. J. Mod. Phys. A3 (1988) 507.

\noindent 16. A.Bilal, Phys. Lett. B227 (1989) 406;31.

\noindent 17. V.A. Fateev and S.L. Lukyanov, Sov. Sci. Rev. A Phys. 15 (1990)1.

\noindent 18. A. Bilal, Introduction to W-Algebras, presented at the Spring
School on St\-ring Theory and Quantum Gravity, Trieste, Italy, 1991, pre\-print
CERN-TH.6083/91.

\noindent 19. P. Bouwknegt and K. Schoutens, Phys. Rep. 223 (1993) 183-276.

\noindent 20. M. Niedermaier,\ Commun.\ Math.\ Phys.\ 148 (1992) 249

\noindent 21. H.R. Karaday\i, Casimir Operators and Eigenvalues for
$A_{_{N-1}}$ Lie algebras, submitted for publications.

\noindent 21. S. Wolfram, ${Mathematica^{TM}}$, Abison-Wesley (1990).
\end